\documentclass[12pt]{article}

\usepackage{epsfig}
\usepackage{cite}
\usepackage{amsmath, amssymb, amsfonts}
\usepackage{color}
\usepackage{latexsym}  
\usepackage{graphicx}
\usepackage{cancel}
\usepackage[colorlinks,bookmarks]{hyperref}
\hypersetup{pdfpagemode=UseNone, pdfstartview=FitH, linkcolor=blue, citecolor=red, urlcolor=blue}

\def\be{\begin{equation}}
\def\ee{\end{equation}}
\def\ba{\begin{eqnarray}}
\def\ea{\end{eqnarray}}

\def\bdm{\begin{displaymath}}
\def\edm{\end{displaymath}}

\def\ga{~\mbox{\raisebox{-.6ex}{$\stackrel{>}{\sim}$}}~}
\def\bq{\begin{quote}}
\def\eq{\end{quote}}

 at 10truept

\newcommand{\p}{\partial}

\newcommand{\Mpl}{M_{\mathrm{Pl}}}

\newcommand{\bea}{\begin{eqnarray}}
\newcommand{\eea}{\end{eqnarray}}

\newcommand{\bi}{\begin{itemize}}
\newcommand{\ei}{\end{itemize}}

\newcommand{\beq}{\begin{equation}}
\newcommand{\eeq}{\end{equation}}
\newcommand{\beqa}{\begin{eqnarray}}
\newcommand{\eeqa}{\end{eqnarray}}
\newcommand{\mpl}{\Mpl}

\def\12{{1 \over 2}}

\def\ltap{\ \raise.3ex\hbox{$<$\kern-.75em\lower1ex\hbox{$\sim$}}\ }
\def\gtap{\ \raise.3ex\hbox{$>$\kern-.75em\lower1ex\hbox{$\sim$}}\ }
\def\gl{\ \raise.5ex\hbox{$>$}\kern-.8em\lower.5ex\hbox{$<$}\ }
\def\roughly#1{\raise.3ex\hbox{$#1$\kern-.75em\lower1ex\hbox{$\sim$}}}

\begin{document}

\thispagestyle{empty}
\begin{flushright}
February 2026
\end{flushright}
\vspace*{1.5cm}
\begin{center}

{\Large \bf Electromagnetic Couplings of Dark Domain Walls} 
 
\vspace*{1.3cm} {\large
Nemanja Kaloper\footnote{\tt
kaloper@physics.ucdavis.edu} }\\
\vspace{.5cm}
{\em QMAP, Department of Physics and Astronomy, University of
California}\\
\vspace{.05cm}
{\em Davis, CA 95616, USA}\\

\vspace{1.5cm} ABSTRACT
\end{center}
We extend Maxwell electrodynamics with a Chern--Simons coupling to a
dark sector top form sourced by domain walls. 
Cosmic birefringence can arise from a distinct mechanism
in which photon polarization is rotated when crossing vacuum interfaces,
rather than through adiabatic propagation in a background field. 
Ultrathin walls induce a finite, frequency-independent polarization
rotation generated by an electromagnetic Chern-Simons term 
localized at the interface. The effect persists even in the absence of
ultralight axions or other propagating scalar degrees of freedom. 
For phenomenologically viable parameters, such walls can generate
cosmic microwave background polarization rotation at the level
$\Delta\vartheta \sim 10^{-3}$ rad, providing a signature of the
topological structure of the dark-sector vacuum.

\vfill \setcounter{page}{0} \setcounter{footnote}{0}

\vspace{1cm}

\newpage

In this work we show that electromagnetism can couple to domain 
walls beyond the Standard Model via a 
Chern--Simons interaction that mixes with a dark top form. 
As a result, bubbles of evanescent dark energy 
\cite{Kaloper:2025goq} may induce detectable polarization 
rotations. This is particularly interesting in light of 
recent hints of cosmic microwave background (CMB) polarization
rotation at the level
\be 
\Delta\vartheta \sim 10^{-3}\,{\rm rad} \, ,
\label{rotang}
\ee
which have generated significant interest as possible evidence
for new dark sector \cite{Komatsu:2022nvu}. The usual interpretation of this effect 
relies on electromagnetic coupling to a light pseudoscalar field, whose slow
spacetime variation induces an adiabatic rotation of photon
polarization which accumulates along the line of sight, 
$\Delta\vartheta \propto \int d x^\mu\,\partial_\mu \phi $ \cite{Huang:1985tt,Harari:1992ea}. 
This is commonly taken to imply that the scalar is extremely light,
$m \lesssim H_{\rm LSS} \sim 10^{-28}\,$eV. Some authors
note that an axion does not need to be as light, 
but \emph{in all cases} they take it to be lighter
than the CMB frequency range, $m < 10^{-4} {\rm eV}$, to rely 
on the adiabatic approximation of \cite{Huang:1985tt,Harari:1992ea}. 

Here we demonstrate that cosmic birefringence can
arise as a \emph{purely interface-localized effect}, generated when
photons cross boundaries between topologically distinct vacua
of a dark sector that support a Chern--Simons
interaction on the wall. This reveals that electromagnetic birefringence falls into
two sharply distinct universality classes:

\begin{itemize}
\item[(i)] \emph{Adiabatic birefringence}, in which polarization
rotation accumulates continuously along the photon trajectory
due to smooth variation of a light axion, considered in past work;

\item[(ii)] \emph{Interface birefringence}, where polarization
rotation changes discretely when a photon crosses a
codimension-one vacuum boundary, and is independent of propagation
history, wall thickness, and photon wavelength below a physical
cutoff.
\end{itemize}

The light axion scenarios considered as sources of cosmic birefringence 
\cite{Huang:1985tt,Harari:1992ea,Ferreira:2023jbu,Takahashi:2020tqv,Kitajima:2023kzu,Nakai:2023zdr}
invariably involve axions lighter than CMB frequencies. Even when localized
interface effects are mentioned, it is argued that such regimes are difficult
to realize or suppressed \cite{Nakai:2023zdr}. 

In sharp contrast, we focus on \emph{ultrathin walls}, which invalidate 
adiabatic description. Instead, the polarization rotation is 
intrinsically a scattering problem across a localized interface (see also \cite{Kaloper:2026ygk}). 
Such walls induce a unitary transformation which rotates photon
polarization by an angle set by the Chern--Simons coupling. Solving
the wave equation, we find that the crossing implements a
finite rotation \cite{Kaloper:2026ygk}
\be 
\Delta\vartheta \simeq \frac{\zeta {\cal Q}}{6{\cal M}^2}\, .
\ee
accompanied by reflection and frequency mixing. In
the weak-coupling limit relevant for observations, reflection is
suppressed and the effect reduces to a pure polarization rotation.

Crucially, this phenomenon persists even in the absence of any
propagating axions. The interaction arises
as a topological remnant of heavy microphysics, encoded in a mixed
Chern--Simons coupling between electromagnetism and a dark top-form
sector. As a result, the effect is sudden rather than adiabatic,
localized at vacuum interfaces, and insensitive to the
cosmological evolution between crossings. 
Our mechanism predicts that cosmic birefringence probes the
topological structure of the dark sector vacuum, rather than the
dynamics of light fields. In particular, a single spherical domain wall, or
a distribution of such walls, can imprint a uniform polarization
rotation across the sky without requiring ultralight particles.

As a concrete realization of interface birefringence, 
we consider the following gauge invariant and Lorentz invariant 
low-energy theory, working in the decoupling limit of gravity, 
$\mpl \to \infty$, and so ignoring spatial curvature 
\cite{Kaloper:2025goq,Kaloper:2025wgn,Kaloper:2025upu}:
\ba
S &\ni& \int d^4 x \Bigl\{-\frac{1}{4 g^2} F_{\mu\nu}^2 - A_\mu J^\mu 
- \frac{\zeta}{4! {\cal M}^2 g^2} {\cal H} 
\epsilon^{\mu\nu\lambda\sigma} F_{\mu\nu} F_{\lambda\sigma} 
- \frac{1}{2} {\cal H}^2 
+ \frac{1}{6} \epsilon^{\mu\nu\lambda\sigma}  
\partial_\mu \bigl( {\cal H} \bigr) {\cal B}_{\nu\lambda\sigma} 
\Bigr\} \nonumber ~~~ \\
&-&{\cal T} \int d^3 \xi \sqrt{ \bigl| \det \left(\eta_{\mu\nu} 
\frac{\p x^\mu}{\p \xi^a} \frac{\p x^\nu}{\p \xi^b} \right) \bigr| } 
- \frac{\cal Q}{6} \int d^3 \xi \, {\cal B}_{\mu\nu\lambda} 
\frac{\p x^\mu}{\p \xi^a} \frac{\p x^\nu}{\p \xi^b} 
\frac{\p x^\lambda}{\p \xi^c} \epsilon^{abc} \, .
\label{cantra}
\ea
Here $A_\mu$ and $F_{\mu\nu}=\partial_\mu A_\nu-\partial_\nu A_\mu$ are the $U(1)$ 
gauge potential and field strength, 
$g$ is the gauge--matter coupling constant, and $J^\mu$ is the conserved current of charged matter 
(normalized to $g^2=1$). The field ${\cal H}$ is a pseudo-scalar magnetic dual of the electric four-form 
(a.k.a. top form) $U(1)$ field strength ${\cal G}_{\mu\nu\lambda\sigma}
=4\partial_{[\mu}{\cal B}_{\nu\lambda\sigma]}$, with 
${\cal B}_{\nu\lambda\sigma}$ the corresponding electric 
three-form gauge potential. The higher-rank forms are 
sourced by membranes of tension ${\cal T}\ge 0$ and charge 
${\cal Q}$. The mass scale ${\cal M}$ originates from the 
ultraviolet completion of the top form sector, and the 
dimensionless parameter $\zeta$ controls the ratio ${\cal Q}/{\cal M}^2$. As discussed in 
\cite{Kaloper:2025goq,Kaloper:2025wgn,Kaloper:2025upu,Luscher:1978rn,Gabadadze:1997kj,Gabadadze:2002ff}, 
${\cal M}$ may be identified with the strong-coupling 
scale of a non-Abelian gauge theory from which the top form 
emerges \cite{Luscher:1978rn}. Below this scale, the naive discrete shift symmetry of the magnetic 
dual ${\cal H} \rightarrow {\cal H} + {\cal Q}$ is 
spontaneously broken by ${\cal H}^2/2$ which is absent in the UV, which is restored by
monodromy branch structure \cite{Witten:1980sp,Witten:1998uka}. 

We imagine that the higher-rank form sector in Eq.~(\ref{cantra}) originates from a 
dark sector, while the familiar 
vector $U(1)$ corresponds to Maxwell electromagnetism. 
The mixing between the two sectors may arise, for example, 
from kinetically mixed very heavy axions -- one in each sector -- monodromized by top forms 
\cite{Dvali:2005an,Dvali:2005zk,Kaloper:2008qs,Kaloper:2008fb,Kaloper:2011jz}. 

A straightforward example of a consistent microscopic completion of the 
effective theory (\ref{cantra}) is provided by a dark sector containing a non-Abelian
 gauge theory that becomes strongly coupled at a very low scale of order 
 milli-eV \cite{Kaloper:2025goq}. As shown by L\"uscher \cite{Luscher:1978rn}, 
 such theories naturally generate topological four-forms. If the four-form is taken to 
 be massive, its longitudinal mode corresponds to 
 a propagating pseudoscalar axion \cite{Dvali:2005an,Dvali:2005zk}.

Crucially, even though our low energy theory 
contains no propagating dark degrees of freedom, electromagnetic couplings 
can still arise consistently. If the emergent dark sector axion is sensitive to 
physics above the dark sector strong coupling scale, it can mix 
either with the Standard Model axion or with a top form sector that solves the 
strong CP problem through non-perturbatively generated operators. This induces a 
suppressed coupling between the dark sector and electromagnetism without introducing 
any light electrically charged particles. Importantly, this interaction survives even if the 
longitudinal axion mass is raised to the dark sector cutoff and the axion is integrated out.

Alternatively, one may also introduce heavy particles charged under the dark gauge group 
and carrying tiny fractional electromagnetic charges, consistent with existing bounds. 
Integrating out these heavy states can again contribute to the effective interaction Eq.~(\ref{cantra}), while 
their radiative corrections in the bulk remain suppressed by large 
masses or non-perturbative effects. Both approaches lead to the resulting low energy 
theory of the form considered here: standard Maxwell electrodynamics 
in the bulk, with the leading new effect given by domain wall localized Chern-Simons terms.

Adopting this logic we treat Eq.~(\ref{cantra}) strictly as a low energy effective field 
theory valid below the dark sector strong coupling scale, after all propagating dark degrees 
of freedom have been integrated out. We do not assume the presence of 
ultralight axions or other light particles coupled to electromagnetism. Instead, the mixed 
electromagnetic-dark Chern-Simons interaction should be understood as a topological 
remnant of heavy microphysics, much like the $\theta$-term in QCD. As a result, conventional 
cosmological constraints associated with light degrees of freedom (such as contributions to $N_{\rm eff}$ 
and large radiative corrections to low energy QED do not apply. The optical activity discussed 
below persists despite the complete absence of light axion fields 
which were crucial in the past proposals. 

For the most part, the theory (\ref{cantra}) behaves as 
conventional Maxwell electrodynamics. Varying the action 
with respect to $A_\mu$ yields
\be
\frac{1}{ g^2 } \partial_\mu F^{\mu\nu}
= J^\nu - \frac{\zeta}{6 {\cal M}^2 g^2 }
\epsilon^{\mu\nu\lambda\sigma} \partial_\mu \Bigl({\cal H} F_{\lambda\sigma}\Bigr) \, .
\label{mmon}
\ee
In the presence of magnetic monopoles, the final term 
would be the CP-violating dyonic contribution to the 
monopole charge \cite{Witten:1979ey}. Varying instead with respect to 
${\cal B}_{\nu\lambda\sigma}$ gives
\be
n^\mu \partial_\mu {\cal H} = {\cal Q} \delta \bigl(r-r(t)\bigr) \, ,
\label{Heq}
\ee
where $r(t)$ describes the wall trajectory, $n^\mu$ 
is the outward-pointing normal vector satisfying 
$n_\mu dx^\mu = dr(t)$, and $r$ is the coordinate 
along $n^\mu$. Away from the walls, ${\cal H}$ is constant and 
hence $\partial_\mu {\cal H}=0$. Consequently, in regions without membranes the final term in 
Eq.~(\ref{mmon}) reduces to $\frac{\zeta {\cal H}}{6 {\cal M}^2 g^2}
\epsilon^{\mu\nu\lambda\sigma}\partial_\mu(F_{\lambda\sigma})$, 
which vanishes identically in Maxwell theory in 
the absence of magnetic monopoles, since $\partial_{[\mu}F_{\lambda\sigma]}=0$. 
One may redefine the field strength as
\be
\tilde F^{\mu\nu} = F^{\mu\nu} + \frac{\zeta {\cal H}}{6 {\cal M}^2}
\epsilon^{\mu\nu\lambda\sigma} F_{\lambda\sigma} \, .
\label{redef}
\ee
Note that for the redefined field strength in (\ref{redef}), in general the field equation for the
dual field strength $^*\tilde F^{\mu\nu} = \frac12 \epsilon^{\mu\nu\lambda\sigma} \tilde F_{\lambda\sigma}$
is not a Bianchi identity even away from the wall. 
Using (\ref{mmon}), (\ref{redef}) and $\partial_{[\mu}F_{\lambda\sigma]}=0$, the relevant field equation is 
$\partial_\sigma \, ^* \tilde F^{\sigma\lambda} = 
- \frac{\zeta g^2}{3M^2} \partial_{\sigma} \bigl({\cal H} F^{\sigma\lambda}\bigr)$,
and so off the wall, $\partial_\sigma \, ^* \tilde F^{\sigma\lambda} = 
- \frac{\zeta {\cal H} g^2}{3M^2} J^{\lambda}$. Hence
in terms of $\tilde F^{\mu\nu}$ all charged particles are dyons \cite{Witten:1979ey}, carrying 
magnetic charges $\mu = -  \sqrt{\frac{\zeta {\cal H}}{3M^2}} g$. Nevertheless, since the ratio of the magnetic 
to electric charges $\mu/g$ is the same for all particles, this is merely a matter of perspective and
can be removed by returning to the original variables $F_{\mu\nu}$, as explained in \cite{Jackson:1998nia}. 

The remaining variation of Eq.~(\ref{cantra}), which will be relevant below, is with respect to ${\cal H}$:
\be
\frac{1}{4!}
\epsilon^{\mu\nu\lambda\sigma} {\cal H}_{\mu\nu\lambda\sigma}
= -{\cal H}
- \frac{\zeta}{4! {\cal M}^2 g^2 }
\epsilon^{\mu\nu\lambda\sigma} F_{\mu\nu} F_{\lambda\sigma} \, ,
\label{topf}
\ee
where ${\cal H}_{\mu\nu\lambda\sigma}=4\partial_{[\mu}{\cal B}_{\nu\lambda\sigma]}$ 
is the spectator electric top form, with its flux fixed by the CP-violating membrane 
sources. Inverting the Hodge dualization in Eq.~(\ref{topf}) gives
\be
{\cal H}_{\mu\nu\lambda\sigma}
= {\cal H}\,\epsilon_{\mu\nu\lambda\sigma}
- \frac{\zeta}{{\cal M}^2 g^2}
F_{[\mu\nu}F_{\lambda\sigma]} \, ,
\label{topform}
\ee
where 
$\epsilon_{\mu\nu\lambda\sigma}\epsilon^{\alpha\beta\gamma\delta} 
F_{\alpha\beta}F_{\gamma\delta}
= -\delta_{\mu\nu\lambda\sigma}^{\alpha\beta\gamma\delta} 
F_{\alpha\beta}F_{\gamma\delta} = -4! F_{[\mu\nu}F_{\lambda\sigma]}$. 
Solving for ${\cal B}$ is straightforward.

The situation becomes more interesting in the presence of walls. 
In the rest frame of the wall, Eq.~(\ref{Heq}) implies 
${\cal H}_-={\cal H}_+-{\cal Q}$, where ${\cal H}_-$ and ${\cal H}_+$ 
are the flux values inside (to the left of) 
and outside (to the right of) the wall. In the
original coordinates, ${\cal H}={\cal H}_-+{\cal Q}\,\Theta \bigl(r-r(t)\bigr)$.
Then to analyze how a membrane wall affects the electromagnetic fields  
we consider walls which are large and treat them as approximately flat.   
Accordingly, we use  
\be
{\cal H} = {\cal H}_- + {\cal Q}\,\Theta \Bigl(n_\mu \bigl(x^\mu - x^\mu_0(t)\bigr)\Bigr) \, ,
\label{step}
\ee
where $x^\mu_0(t)$ is a parametric specification 
of the wall location and $n_\mu$ is the outward normal.

We now examine the electromagnetic fields along a 
line of sight that crosses the wall. This problem shares
some common features with the propagation of light across axion 
domain walls \cite{Huang:1985tt,Harari:1992ea}
but there are some important differences. 
As in \cite{Huang:1985tt,Harari:1992ea}, we find
that variables $\hat F_{\mu\nu} = F_{\mu\nu} + \frac{\zeta}{12 {\cal M}^2} 
\epsilon_{\mu\nu\lambda\sigma} {\cal H} F^{\lambda\sigma}$ introduced in Eq.~(\ref{redef}) 
are convenient for understanding how electromagnetic fields cross the wall.
There is an important distinction. In \cite{Huang:1985tt,Harari:1992ea} 
they consider walls which are thick compared to the 
wavelength of the incident electromagnetic waves
and use adiabatic limit to describe the propagation. In contrast, in our case the walls
are ultra-thin compared to light wavelengths and so we must use the complementary ``sudden" approximation. 
Conceptually, this is like Fresnel's description of light refraction on interfaces between 
dielectrics \cite{Jackson:1998nia}. However there are important simplifications. 

First off, having infinite walls, 
and going to their rest frame, we pick the coordinates so wall sits at $z=0$
plane. Then using the boost invariance of the wall in 
the $z-t$ plane  \cite{Huang:1985tt,Harari:1992ea}, we can go 
to the frame where the incident waves come along the normal to the wall. Finally, since ${\cal H}$ 
off the wall is constant and we are ignoring the curvature of space, the dispersion relation 
is $\omega^2 = \vec k^2$ on both sides, and the index of refraction is $n_{\pm} = 1$.  
So, the waves are impacting the wall at right angles, 
propagating both after and before as waves in vacuum. 
Any specific case where the wave vector and wall 
normal are at an angle can be reduced to this by a Lorentz transformation. 

Next, since the wall is ultra-thin, the derivatives of ${\cal H}$ are large but integrable,
and ultra-localized. Since the field equations (\ref{mmon}) are  
$\partial_\mu \tilde F^{\mu\nu} = 0$, to them 
and the variables $\tilde F^{\mu\nu}$ the large derivatives
of ${\cal H}$ are invisible. The problem is with $F_{\mu\nu}$ and the Bianchi 
identities $\partial_{[\mu} F_{\nu\lambda]} = 0$, which 
are sensitive to the large derivatives on the wall. The variables 
$\hat F^{\mu\nu}$ balance this out, averaging the field variables at the wall and replacing them with
principal values, which preserves the $(+) \leftrightarrow (-)$ symmetry across the wall. 
 
To understand what happens, we switch to 
$E^i=F^{0i}$ and $B^i = \frac12 \epsilon^{ijk} F_{jk}$; the redefined 
average gauge field strengths are, using $\sigma=\frac{\zeta {\cal H}}{6{\cal M}^2}$ 
\be
\vec {\hat E} = \vec E - \sigma \vec B \, , 
\qquad \qquad
\vec {\hat B} = \vec B + \sigma \vec E \, .
\label{ebfieldsdefs}
\ee
The full set of Maxwell equations with 
vanishing local charges and currents is
\ba
&&\vec \nabla \cdot \vec {\hat E}  =  \vec \nabla \cdot \bigl( \sigma \vec B  \bigr)  \, ,  \qquad ~~~\,
\vec \nabla \times \vec {\hat B} - \partial_t \vec {\hat E} + \vec \nabla \times \bigl( \sigma \vec E \bigr) 
+ \partial_t  \bigl(\sigma \vec B \bigr) = 0 \, , 
\label{maxeqs1}\\
&&\vec \nabla \cdot \vec {\hat B} = \vec \nabla \cdot \bigl( \sigma \vec E  \bigr) \, , 
\qquad ~~~\,
\vec \nabla \times \vec {\hat E} + \partial_t \vec {\hat B} + \vec \nabla \times \bigl( \sigma \vec B \bigr) 
- \partial_t  \bigl(\sigma \vec E \bigr) = 0 \, .
\label{maxeqs2}
\ea
The first two (\ref{maxeqs1}) are $\partial_\mu \tilde F^{\mu\nu} = 0$, and the second two 
(\ref{maxeqs2}) are the Bianchi identities $\partial_{[\mu} F_{\nu\lambda]} = 0$ after
we added and subtracted terms $\propto$ $\vec \nabla \cdot ( \sigma \vec E)$, 
$\vec \nabla \times ( \sigma \vec B )$ and 
$\partial_t  (\sigma \vec E )$, and used (\ref{ebfieldsdefs}) under derivatives. We now take $\sigma \ll 1$ 
and treat the wall interactions as a perturbation of the vacuum equations, motivated by 
the case of axion walls \cite{Huang:1985tt,Harari:1992ea}.

We stress again that although our setup may appear reminiscent of optical 
activity processes in axionic domain walls \cite{Huang:1985tt,Harari:1992ea}, 
the underlying physical mechanism is completely different. 
In axion domain wall scenarios, optical activity arises from long-range 
axion field variations which require the axion to be 
light enough for the effect to be operative at 
CMB frequencies \cite{Huang:1985tt,Harari:1992ea,Ferreira:2023jbu}.
In stark contrast, here there are no propagating axion-like degrees of freedom at all. 
The optical activity arises from a purely topological Chern-Simons interaction localized on the domain wall, 
and persists even when all axion-like fields have been integrated out. 

Since for $\vec E, \vec B$ in the absence of charges and currents
$\rho = \vec j = 0$ we have $\vec \nabla \cdot \vec E = 0$ 
and $\vec \nabla \times \vec B = \partial_t \vec E$ to leading order in the $\sigma$ expansion,
and $\partial_t \sigma = 0$, $\vec \nabla \sigma  \parallel \vec n \parallel \vec k$, where $\vec k$ is the 
wave vector, and the waves are right handed triads $\vec k, \vec E, \vec B$, the
Eqs (\ref{maxeqs1}), (\ref{maxeqs2}) reduce to
\be
\vec \nabla \cdot \vec {\hat E}  = 0  \, , \qquad 
\vec \nabla \times \vec {\hat B} - \partial_t \vec {\hat E} = - \vec \nabla \sigma \times \vec E 
\, , \qquad  
\vec \nabla \cdot \vec {\hat B} = 0 \, , 
\qquad 
\vec \nabla \times \vec {\hat E} + \partial_t \vec {\hat B}  
= - \vec \nabla \sigma  \times \vec B \, .
\label{maxeqs20}
\ee        
We integrate Eqs.~(\ref{maxeqs20}) over Gaussian 
pillboxes encroaching the wall, with bases parallel to it. 
The first and the third give the answers for $\vec {\hat E}, \vec {\hat B}$ which look just like the 
standard results for $\vec E, \vec B$ across the charge- and current-free interfaces between  
dielectrics since those equations look the same \cite{Jackson:1998nia}. The second and the fourth equations 
involve an extra term $\propto \vec \nabla \sigma$ and 
we need its circulation along a closed infinitesimal
rectangle straddling the interface. 
Using $\vec \nabla \sigma \parallel \vec n$, in both cases the contributions
from the paths parallel with the interface vanish since $\vec \nabla \sigma = 0$
 on either side, and those on the paths crossing the interface are orthogonal to
the paths. So the boundary conditions controlling the fields 
across the wall are 
\be
\vec n \cdot \Bigl(\vec {\hat E}_+- \vec {\hat E}_- \Bigr)  = 0  \, , ~~~ 
\vec n \times \Bigl( \vec {\hat B}_+-  \vec {\hat B}_- \Bigr) = 0 \, , ~~~  
\vec n \cdot \Bigl(\vec {\hat B}_+- \vec {\hat B}_- \Bigr)  = 0 \, , 
~~~  
\vec n \times \Bigl( \vec {\hat E}_+-  \vec {\hat E}_- \Bigr) = 0  \, .
\label{maxeqsbcs}
\ee
These look precisely the same as the standard boundary conditions on
dielectric interfaces -- but for  $\vec {\hat E}, \vec {\hat B}$. Ergo, it is these variables
that cross the wall `smoothly'. In turn, the canonically normalized fields
undergo a propagation induced {\it electromagnetic duality 
transformation} of the original monopole-free $U(1)$ field strengths 
\cite{Jackson:1998nia,Shapere:1991ta} on the wall. 
Defining the rotation angle $\vartheta$ via
$\cos\vartheta = (1+\sigma^2)^{-1/2}$ and $\sin\vartheta = \sigma(1+\sigma^2)^{-1/2}$, 
so that $\sigma=\tan\vartheta$, and factorizing the transformation encoded by (\ref{maxeqsbcs}) 
into a rotation matrix and a rescaling we find 
\be
\frac{1}{g^2 \cos\vartheta_-}
\begin{pmatrix}
\cos\vartheta_- & -\sin\vartheta_- \\
\sin\vartheta_- & \cos\vartheta_-
\end{pmatrix}
\begin{pmatrix}
\vec E_{-} \\
\vec B_{-}
\end{pmatrix}
=
\frac{1}{g^2 \cos\vartheta_+}
\begin{pmatrix}
\cos\vartheta_+ & -\sin\vartheta_+ \\
\sin\vartheta_+ & \cos\vartheta_+
\end{pmatrix}
\begin{pmatrix}
\vec E_{+} \\
\vec B_{+}
\end{pmatrix} \, .
\label{EBmatrixa}
\ee
The resulting optical activity is the field-space rotation 
\cite{Jackson:1998nia,Shapere:1991ta} 
\be
{\cal R}(+\rightarrow -) =
\begin{pmatrix}
\cos\vartheta_- & \sin\vartheta_- \\
-\sin\vartheta_- & \cos\vartheta_-
\end{pmatrix}
\begin{pmatrix}
\cos\vartheta_+ & -\sin\vartheta_+ \\
\sin\vartheta_+ & \cos\vartheta_+
\end{pmatrix}
=
\begin{pmatrix}
\cos\Delta \vartheta & - \sin \Delta \vartheta  \\
\sin \Delta \vartheta &  \cos \Delta \vartheta
\end{pmatrix}
\, ,
\ee
which strictly speaking in our case 
holds only for small $\Delta \vartheta$ as we neglected ${\cal O}({\cal \sigma}^2)$ 
contributions in deducing (\ref{maxeqsbcs}). 
Specifically, if $\vec E$ field initially points along the $x$ direction and 
$\vec B$ field along $y$ outside the wall, then after crossing  
both are rotated by the same frequency independent angle,
\be
\Delta\vartheta \simeq \frac{\zeta {\cal Q}}{6{\cal M}^2} \, .
\label{polangle}
\ee
Curiously, the transformation in Eq.~(\ref{EBmatrixa}) also suggests a rescaling 
of electric and magnetic fields across the wall.
However, for small $\sigma$ this effect enters only at subleading order ${\cal O}(\sigma^2)$. 
In a complete description  
\cite{Kaloper:2026ygk} this rescaling is countermanded by unitarity,
since the subleading reflection from the wall, neglected here since it appears
at higher orders in perturbation theory, compensates these terms. 
By contrast, the polarization rotation arises already at linear order 
and is robust. We therefore focus on it as the leading physical signature 
of the interface interaction. 

Any bounds on the electromagnetic couplings to the walls depend on 
encountering walls, which is controlled by the nucleation rates of membranes charged 
under ${\cal H}$. In the toy model of thin walls coupled to electromagnetism in 
Eq.~(\ref{cantra}), the nucleation rates depend  
on the background field values and theory parameters. 
A crucial subtlety arises from the UV behavior of the theory. 
If the top form description remains valid up to a high cutoff, so  
that the membrane radius at nucleation exceeds the inverse cutoff, 
the semiclassical description of nucleation is reliable. In this case the rate can be 
computed using standard techniques \cite{Coleman:1977py,Callan:1977pt,Garriga:1993fh}, 
treating the process as a Schwinger-like discharge of 
extended objects \cite{Schwinger:1951nm,Brown:1987dd,Brown:1988kg}. 

By contrast, if the top form emerges only below a symmetry-breaking scale -- as in the case of 
QCD \cite{Luscher:1978rn,Gabadadze:1997kj,Gabadadze:2002ff} 
or in discretely evanescent dark energy proposal \cite{Kaloper:2025goq} -- 
symmetry restoration above that scale renders vacua degenerate 
and removes the top form from the spectrum. So if the membrane size at 
nucleation were shorter than this scale, a full UV completion is 
required\footnote{A simple way to understand this is the Heisenberg uncertainty principle 
$\Delta x \Delta p \ga 1$, which says that if we 
cut off the momenta at scales $p_* \sim \Lambda$, we cannot 
describe phenomena of spatial extent less than $\Delta x \sim 1/\Lambda$.}. 
In our examples, the UV completion is provided by asymptotically free 
Yang--Mills theories with chiral symmetry and degenerate vacua in the ultraviolet. 
In this regime there is neither a top form to discharge nor an 
energy difference between vacua. They only emerge in the IR. 
Therefore the nucleation rate must be extremely suppressed. 

To proceed, we rewrite Eq.~(\ref{cantra}) in Euclidean space 
\cite{Kaloper:2022oqv,Kaloper:2022utc}. 
Defining the Euclidean action via $iS=-S_E$, 
and integrating the bilinear term by parts to account for 
membrane boundary conditions, we obtain
\ba
S_E &=& \int d^4x \Bigl(
\frac{\zeta}{6 {\cal M}^2 g^2} {\cal H}\,\vec E \cdot \vec B
+ \frac12 {\cal H}^2
+ \frac{1}{6} \epsilon^{\mu\nu\lambda\sigma}
\partial_\mu({\cal H}) {\cal B}_{\nu\lambda\sigma}
\Bigr)
\nonumber \\
&+& {\cal T} \int d^3 \xi \sqrt{\gamma}_{\cal B}
- \frac{{\cal Q}}{6} \int d^3 \xi \,
{\cal B}_{\mu\nu\lambda}
\frac{\p x^\mu}{\p \xi^\alpha}
\frac{\p x^\nu}{\p \xi^\beta}
\frac{\p x^\lambda}{\p \xi^\gamma}
\epsilon^{\alpha\beta\gamma} \, ,
\label{euclid}
\ea
with the standard Lorentzian-to-Euclidean mapping \cite{Kaloper:2022oqv,Kaloper:2022utc}. 
We retain the background contribution $\vec E\cdot\vec B/g^2$. When the 
fields are approximately constant, the dominant contribution arises from $O(4)$-symmetric 
tunneling; subleading channels yield at most ${\cal O}(1)$ corrections and will be neglected. 
Treating $\vec E\cdot\vec B/g^2$ as stationary and constant over the region 
relevant for the tunneling is a good approximation for small nucleated bubbles. 

Using (\ref{euclid}), the tunneling configurations correspond 
to slices of a four-dimensional sphere $S^4$
glued along a fixed $S^3$ latitude. These configurations describe 
Euclidean world-volumes of spherical membranes
with tension ${\cal T}$ and charge ${\cal Q}$. Detailed 
analyses of such processes can be found in
\cite{Brown:1987dd,Brown:1988kg,Kaloper:2022oqv,Kaloper:2022utc}.
In the limit $\Mpl \to \infty$, the only relevant configuration is the Euclidean bounce
connecting backgrounds with vacuum energy
\be
V = \frac{1}{2} {\cal H}^2 + \frac{\zeta}{6{\cal M}^2 g^2}\, {\cal H}\,\vec E \cdot \vec B \, ,
\label{finpots}
\ee
where the interior and exterior values of ${\cal H}$ differ by a single unit of charge,
$\Delta {\cal H} = {\cal Q}$, as implied by the Euclideanized equation (\ref{Heq}).
At least the initial vacuum energy must be non-negative.
The corresponding energy difference $\Delta V$ inside 
the membrane, arising from the flux discharge,
must balance the cost of creating a membrane 
with tension ${\cal T}$ \cite{Coleman:1977py}.

The ``statics'' of the configuration are governed by d'Alembert's principle of virtual work,
which compares the action (\ref{euclid}) evaluated on a configuration 
containing a single bubble to that of the smooth background.
For constant $\Delta V$ and ${\cal T}$, the relevant volume factors are
$V_{S^4} = \pi^2 r_0^4/2$ and $V_{S^3} = 2\pi^2 r_0^3$, yielding
\be
S_{\rm membrane}
= 2\pi^2 r_0^3 {\cal T} - \frac{1}{2}\pi^2 r_0^4 \Delta V \, ,
\label{sbounce}
\ee
where, from (\ref{finpots}),
\be
\Delta V \simeq
\Bigl( {\cal H} + \frac{\zeta}{6{\cal M}^2 g^2}\,\vec E \cdot \vec B \Bigr) {\cal Q} \, .
\label{delv}
\ee
Minimizing (\ref{sbounce}) with respect to $r_0$ gives
\be
r_0 = \frac{3{\cal T}}{\Delta V} \, , \qquad\qquad
B = \frac{\pi^2}{2}\,{\cal T}\,r_0^3 \, .
\label{bounce}
\ee
The nucleation rate per unit spacetime volume is $\Gamma = A e^{-B}$
\cite{Coleman:1977py,Callan:1977pt}.
In the limit $\Mpl \to \infty$, the prefactor $A$ 
was computed in \cite{Garriga:1993fh}, yielding
\be
\Gamma \simeq {\cal T}^2 r_0^2
\exp \Bigl(-\frac{\pi^2}{2}\,{\cal T}\,r_0^3\Bigr) \, .
\label{nucrate}
\ee
Requiring $r_0>0$ implies $\Delta V>0$. From (\ref{delv}), two regimes arise:
(i) $\vec E \cdot \vec B/g^2 \ll 3{\cal M}^2{\cal H}/\zeta$, 
and (ii) $\vec E \cdot \vec B/g^2 \gg 3{\cal M}^2{\cal H}/\zeta$. 

In the first regime, the bounds reduce to the familiar 
cosmological constraints from the CMB,
dating back to \cite{Zeldovich:1974uw} for large, horizon-crossing domain walls.
Recently it has been suggested that additional 
bounds may arise if the bubble-sector scales
are very low and nucleations become prolific when 
late-time vacuum energy dominates \cite{Koren:2025ymq}.
Both situations have been explored in the context of relaxing strong CP violation
\cite{Kaloper:2025wgn,Kaloper:2025upu} and discretely evanescent dark energy
\cite{Kaloper:2025goq}. Such phenomena, including possible gravitational-wave 
signatures, could provide additional observational probes of these dynamics.

We now estimate how the nucleation rate is affected by the electromagnetic
$\vec E \cdot \vec B/g^2$ term.
The most controlled laboratory scenario for 
enhancing nucleation could involve stationary, uniform,
and parallel electric and magnetic fields.
In this case, the magnitude of $\vec E \cdot \vec B/g^2$ 
is limited by the maximum steady fields
achievable in practice.
Laboratory magnetic fields do not exceed $\sim{\cal O}(10)\,{\rm Tesla}$,
while pulsed lasers can produce electric fields as large as
$\sim 10^{13} -  10^{14}\,{\rm V/m}$, implying
$\vec E \cdot \vec B/g^2 \simeq 0.1\,({\rm keV})^4$.

When ${\cal M}^2{\cal Q}/\zeta \gg 0.1\,({\rm keV})^4$,
the electromagnetic correction to the bubble production rate is negligible.
The nucleation rate is then suppressed by the vacuum tunneling barrier, 
similar to the previous case, 
and the parameter ranges relevant for restoring strong CP violation
\cite{Kaloper:2025wgn,Kaloper:2025upu} remain valid.
In that example, $\sqrt{\cal Q}\sim{\cal M}\sim 3\,{\rm keV}$ and $\zeta\lesssim 1$,
dominating over $\vec E \cdot \vec B/g^2$, 
which implies that the electromagnetic coupling of the walls can be substantial,
with the coefficient of $\vec E \cdot \vec B/g^2$ in (\ref{cantra})
being close to unity.

A key difference relative to cosmological bubble production is that the relevant
four-volume is not the Hubble volume but the ``world-volume of the experiment,''
$\Gamma\,\Omega_{\rm lab}$, adapting the approach of
\cite{Guth:1982pn,Turner:1992tz}.
Here $\Omega_{\rm lab}$ is the spacetime volume occupied by the region with
nonzero $\vec E \cdot \vec B/g^2$ during the experiment.
For steady fields probed with laser pulses, a generous estimate gives
$\Omega_{\rm lab} \sim ({\rm m})^3 \times 10^{-18}\,{\rm s}
\sim 10^{12}\,{\rm eV}^{-4}$. 

Putting it together, if the membranes responsible for CP restoration
\cite{Kaloper:2025wgn,Kaloper:2025upu} couple to electromagnetism as proposed here,
and if their cosmological production rate satisfies
$\Gamma \sim H_{\tt QCD}^4 \sim 10^{-36}\,{\rm eV}^4$
so that all discharges complete before BBN \cite{Kaloper:2025wgn,Kaloper:2025upu},
then the expected laboratory event rate is
$\Gamma\Omega_{\rm lab} \sim 10^{-24}$ for $\zeta \sim{\cal O}(1)$.
Further reducing the electromagnetic coupling suppresses this rate even more.
The bubble radius at nucleation in this regime is
$r_0 \sim {\cal T}/\Delta V \sim {\rm few}\,{\rm keV}^{-1}$,
consistent with the experimental energy scales and the cutoff of the 
theory, which validates our estimate. In sum, the nucleation rate is extremely suppressed. 

If instead the membrane tension and charge are much smaller,
as in the dark energy model of \cite{Kaloper:2025goq},
the semiclassical barrier suppression is absent unless the electromagnetic coupling
is extremely small, $\zeta \lesssim 10^{-23}$.
In this case, the semiclassical nucleation radius would be
$r_0 \sim {\cal T}/\Delta V \sim 10^{-21}\,{\rm eV}^{-1}$,
requiring the effective description to remain valid up to energies
${\cal E}\gtrsim 10^{12}\,{\rm GeV}$. However, the cutoff of this sector is 
only of order milli-eV \cite{Kaloper:2025goq}.
Above the cutoff, the dark sector becomes an asymptotically free Yang--Mills theory
with degenerate vacua, and no top form. 
Thus steady lab fields will not yield prolific nucleation rates.

Much stronger but highly localized electromagnetic fields occur in particle colliders
and have been studied in connection with CP violation and the chiral magnetic effect
\cite{Fukushima:2008xe}.
At the RHIC and LHC, heavy-ion collisions can generate magnetic fields
$\sim10^{14} - 10^{15}\,{\rm Tesla}$ and electric fields
$\sim10^{18} - 10^{19}\,{\rm V/m}$.
Because the ions are ultra-relativistic, 
the fields satisfy $|\vec E| \sim |\vec B|$.
While individual ions primarily generate radiative fields with
$\vec E \cdot \vec B \ll \vec B^2$,
stochastic fluctuations in the swarm of ${\cal O}(100)$ ions produce peak regions
with $\vec E \cdot \vec B \sim \vec B^2$.
This yields $\vec E \cdot \vec B/g^2 \lesssim 10^{30}\,{\rm eV}^4$
\cite{Bzdak:2011yy,Deng:2012pc},
but these fields persist only over spatiotemporal scales
$\sim10^{-16}\,{\rm m}\sim10^{-10}\,{\rm eV}^{-1}$,
so that $\Omega_{\rm lab}\sim10^{-40}\,{\rm eV}^{-4}$.

When $\Delta V$ is dominated by $\vec E \cdot \vec B/g^2$ in (\ref{delv}),
the semiclassical bubble radius at nucleation 
$r_0\sim{\cal T}/\Delta V$ will be small, and the semiclassical 
barrier is absent. For the discretely evanescent dark energy model of \cite{Kaloper:2025goq},
with ${\cal M}\sim{\cal T}^{1/3}\sim{\cal Q}^{1/2}\sim10^{-3}\,{\rm eV}$,
one finds $\Delta V\sim\zeta \times10^{30}\,{\rm eV}^4$ and
$r_0\sim10^{-39}\,{\rm eV}^{-1}/\zeta$. 
This is far above any reasonable UV cutoff
of the effective theory. The IR description fails, but 
since the UV completion has degenerate vacua and no top forms, 
the nucleations are extremely suppressed. 

For the QCD-motivated scenario of \cite{Kaloper:2025wgn,Kaloper:2025upu},
the bubble radius is larger by the cube of the ratio of cutoffs,
$r_0\sim10^{-21}\,{\rm eV}^{-1}$,
corresponding to energies ${\cal E}\sim10^{12}\,{\rm GeV}$.
If the discharging sector in this case is UV completed by an 
asymptotically free gauge theory which enters strong coupling
and chiral symmetry breaking below $10^{12} {\rm ~GeV}$, below which
the top form is induced, the membrane production in these circumstances 
would also evade copious production in heavy-ion collisions,
despite large electromagnetic fields and 
significant wall couplings to electromagnetism at low energies.
A similar conclusion applies to magnetars, whose enormous electromagnetic fields
also involve energies where top forms and membranes have already decoupled.

Additional constraints from cosmology concern a straggler bubble 
nucleated far from the peak nucleation epoch could survive
percolation and later reenter our Hubble volume.
If its tension were too large, the resulting wall would distort the CMB.
Requiring that its total energy be smaller than the energy in a Hubble volume,
$E_{\rm hubble}\simeq\rho_0/H_0^3$, by a factor of $10^{-6} - 10^{-7}$
\cite{Zeldovich:1974uw} implies
${\cal T} < 10^{-6}\Mpl^2 H_0 \sim {\rm few}\times(100\,{\rm keV})^3$. 

The membranes considered in
\cite{Kaloper:2025goq,Kaloper:2025wgn,Kaloper:2025upu}
can easily satisfy both this bound and the limits derived from the absence of 
nucleations in strong electromagnetic fields, while still coupling nontrivially
to electromagnetism via the Maxwell--Chern--Simons term in (\ref{cantra}).
This is particularly intriguing in light of claims that weak optical activity 
might have been observed \cite{Komatsu:2022nvu}, at the level of, roughly, 
\be
\Delta \vartheta \sim 10^{-3}\,{\rm radians} \, , 
\label{variations}
\ee
based on CMB polarization surveys. 
Such effects may easily arise from 
electromagnetic interactions with an evanescent dark-energy domain wall,
with an electromagnetic Chern-Simons coupling $\zeta{\cal Q}/(6{\cal M}^2)\sim{\rm few}\times10^{-3}$. 
Note that in our cases the optical activity arises purely due to the topological features of the
dark sector gauge theories, without requiring ultralight axions, distinguishing it from 
standard realizations in the literature 
\cite{Huang:1985tt,Harari:1992ea,Ferreira:2023jbu,Takahashi:2020tqv,Kitajima:2023kzu,Nakai:2023zdr}. 
In the discretely evanescent dark-energy scenario,
this would also be accompanied by an ${\cal O}(1)$ jump in the dark-energy density at low redshift.
These coincidences merit further scrutiny. 

\vskip.5cm

{\bf Acknowledgments}: We thank G. D'Amico and A. Westphal for discussions. 
This research was supported in part by the DOE Grant DE-SC0009999.

\end{document}